%Paper: hep-ph/9205246
%From: GOLDSTEI@pearl.tufts.edu
%Date: Fri, 29 May 1992 22:54 EDT
%Date (revised): Sat, 30 May 1992 16:59 EDT

% get my favourite font/magnification 1-5 available
\font\sanss=cmss10
%\font\sansb=cmssdc10
\font\sansb=cmssbx10

\magnification=\magstep1
\sanss
%----------------------------------------------
\baselineskip=15pt
\parindent=20pt
\noindent
\voffset=0.0truecm
\raggedbottom
\nopagenumbers

\def\center{\centerline}
\def\np{\vfill\eject}
\settabs 8\columns
\bigskip
\noindent
\parskip=0.20cm
\parindent=13pt
\hsize=17.truecm
\vsize=23.5truecm
\hoffset=0.0truein
\voffset=0.0truein
\raggedbottom
\nopagenumbers

\def\center{\centerline}
\def\np{\vfill\eject}
% ----------------------------------------------
%     my definitions for the symbols used
% ----------------------------------------------

%-----------------------------------------------
\settabs 13\columns
\baselineskip=15pt
\+&&&&&&&TUFTS-TH-92-G01\cr
\+&&&&&&&hep-ph/9205246\cr
\+&&&&&&&&&&&May 15, 1992\cr
\null\vskip 2.5 truecm
\center {ON OBSERVING TOP QUARK PRODUCTION AT THE TEVATRON}
\baselineskip=20pt
\+\cr
\+\cr
\center { Gary R. Goldstein and K. Sliwa}
\smallskip
\baselineskip=10pt
\center { Department of Physics}
\center { Tufts University}
\center { Medford, Massachusetts 02155 USA}
\bigskip
\baselineskip=20pt
\center { R.H. Dalitz}
\smallskip
\baselineskip=10pt
\center { Department of Theoretical Physics}
\center { University of Oxford}
\center { 1 Keble Road, Oxford OX1 3NP UK}
\bigskip
\vskip 2.0 truecm
{\sansb ABSTRACT~~}
\sanss
\smallskip
A technique for separating top quark production from Standard
Model background events is introduced. It is applicable to
the channel in which one top quark decays semi-leptonically and its
anti-quark decays hadronically into three jets, or vice versa.
The method is shown
to discriminate dramatically between Monte Carlo generated events
with and without simulated top quarks of mass around 120 GeV and above. The
simulations were performed with CDF detector characteristics
incorporated, showing that the method is applicable to existing data.
\np
     With the discoveries of the tau lepton and then of the
bottom quark, it became clear that the completion of a
third generation in the Standard Model required a sixth quark flavor, now
named "top". The successes of this model are now so
numerous and so detailed that few doubt the existence of the top quark[1],
yet its discovery has remained elusive. The only firm knowledge that we
have about its mass is that it exceeds 89 GeV[2], a bound deduced from the
empirical di-lepton inclusive total cross-section corresponding to top-
antitop production followed by decays which include a lepton. Some
theoretical arguments suggest that its mass may not be larger than about
200 GeV, and probably less. Unless its mass is much greater than this, top
-antitop pairs should be produced by the Tevatron collider, but
formidible backgrounds have made its detection
quite difficult.

     Owing to its uncommonly large mass, top decay is expected to have some
unique features[3,4]. The most important of these are that decay through
the (V-A)
weak interaction, to a b-quark and a real W boson will dominate, and that,if
the top quark mass m$_t$ is above about 120 GeV, this weak decay will occur
more
rapidly than hadronization[3].

    Recently, two of us [3] developed a scheme for analysing top-antitop
pair production in proton-antiproton annihilation through their leptonic decay
modes. The two-step decay t$\to$bW$^{+}$, W$^{+}$$\to l\nu$, led us to
construct a paraboloid
surface of possible top three-momenta, based on the momenta of the b-jet and
the lepton. Each possible top energy corresponds to a circular cross-section
of it; each possible top mass to an elliptic section of it.

     A $t \bar{t}$ pair is produced by the interaction of a parton of the
proton
with an antiparton of the antiproton. It is known empirically that the partons
have quite limited transverse momentum. Hence the simplest parton model
requires that the net transverse momentum of the produced pair be zero;
however,
to leading order in QCD, some transverse momentum can arise from gluon
bremsstrahlung or the emission of hard gluons. The result[3] is that a real
$t \bar{t}$ pair corresponds to points on the transverse plane projection of
their
two ellipses, such that there is near cancellation of their transverse momenta.
This criterion has been used for the analysis of the CDF $t \bar{t}$ candidate
event
[5], with the conclusion that the mass m$_{t}$ must be about 131 GeV[6], if
this
identification of the event is correct. Although this event was very striking,
and well-fitted in this way, this analysis was not enough to prove that this
event was an example of $t \bar{t}$ pair production.

     The W also decays hadronically, through the two quark channel,
more probably than leptonically (by a factor of about 6 to 1 for
the number of $q-\bar{q}$  combinations versus any single lepton
channel). Hence a larger signal for top pair production is the
event configuration
$$\eqalignno{
p+\bar{p}&\to t+\bar{t}+X&(1)\cr
t        &\to \bar{l}+\nu + (b-jet)\cr
\bar{t}  &\to (q-jet) + (\bar{q}-jet) + (\bar{b}-jet)\cr}$$
\noindent
in which there is a single very energetic lepton (from the W
emitted by one of the top quarks) accompanied by 4 energetic jets
(3 from the decay of the other top quark and one accompanying the
lepton).

     For high mass top quarks the cross section for this channel,
(1), is expected to stand out over the anticipated background[7].
which is approximated by the production of a W boson along with
multiple jets originating from hard gluon bremsstrahlung and quark
antiquark pair production. When felicitous variables are chosen,
the separation can be quite pronounced, and that is the major result of
this paper. In the following we will show that for events satisfying
certain kinematic
criteria, a probabilistically weighted mass distribution function,
to be referred to as the accumulated probability distribution, can be
defined that will distinguish
$t\bar{t}$ events from Standard Model background.

     In an event like (1) the 4-momenta of the three jets
associated with the hadronic decay will determine a total
4-momentum for the hadronically decaying top; both the momentum
$\vec p_t$ and the mass m$_{t}$*  will be determined. The lepton and remaining
b-jet 4-momenta, along with the mass m$_{t}$* just determined, define an
ellipse of possible 3-momenta $\vec p_{t'}$ for the other top quark, as
already shown[3]. What is the relation between $\vec p_t$ and the
ellipse for the presumed mass?

     The hypothesis that the production mechanism involved limited
transverse momentum leads to the requirement that the resulting
transverse momentum projection ($\vec p_t$)$_{T}$+($\vec p_{t'}$)$_T$ be near
zero. Then the
negative of the transverse vector -($\vec p_t$)$_T$ will be near points on the
ellipse, as illustrated in Fig.1. This near coincidence severely
restricts the kinematics of top production events. The combination
of the existence of an ellipse for m$_{t}$* and a near cancellation of
transverse momenta consitutes our kinematic fitting criteria. A
kinematic configuration satisfying those criteria will constitute
a "match" to the criteria.

     The probabilistic weighting for any particular "matched" kinematic
configuration for process (1) depends on several factors.
The (normalized) leptonic decay probability depends on
the top mass
and the lepton and b-jet 4-momenta through the relation
$$\eqalignno{P_l(b,l)&={
{2(2b \cdot l - M_W^2)(m_t^2-m_b^2+M_W^2-2b \cdot l)}
\over{(m_t^2-m_b^2)^2+M_W^2(m_t^2+M_b^2)-2M_W^4}}&(2)}$$
\noindent
where $b$ and $l$ are the b-jet and lepton 4-momenta. The hadronic
decay probability can be written only when the flavors of the three
quarks or their jet fragmentation is known, a circumstance that
will be difficult to sort out experimentally (hence we will simply
take 1 for this factor in the calculations that follow). The total
transverse momentum will be distributed normally about zero with a
width on the order of 10 \% of the produced mass[8]. For subsequent
calculations we will take a step function of width 0.1m$_{t}$ for
transverse momentum, for simplicity.

     The production mechanism favored for any particular kinematic
match depends on the Bjorken x values for the initiating partons,
that is the structure functions F$_1$(x$_1$) and F$_2$(x$_2$), and
the parton
subprocess center-of-mass energy and momentum transfer through the
cross section. The relative probability for producing the
kinematical configuration is then given by
$$\eqalignno{ P_{x_1,x_2}&=
{{
\sum_{i=qq,gg} F_i(x_1)F_i(x_2)
     {
        {d \sigma}
        \over
        {d \hat{t}}
     }
(\hat{s},\hat{t})_i
}
\over
{
\sum_{i=qq,gg}
     {
        {d \sigma}
        \over
        {d \hat{t}}
     }
(\hat{s},\hat{t})_i
}}
&(3)\cr}$$
\noindent
where the relevant variables are obtained from the energies and
momenta of the top quarks in the $p \bar{p}$ lab frame,
\medskip
\+&&$x_{1,2}=(E_t-E_{t'}\pm
(p_{tL}+p_{t'L}))/2P$;\cr
\smallskip
\+&&$\hat{s}=x_1 x_2 s;$\cr
\smallskip
\+&&$\hat{t}=m_t^2-x_1 \sqrt(s) (E_t-p_{tL})$\cr
\smallskip
\noindent
with $P$ the proton momentum, s the square of the $p\bar {p}$
center-of-mass energy, and $p_{tL}$ the top longitudinal momentum.
The {\it a priori\/} probability for a top quark production through this
channel (1) with a given kinematical configuration will be the
product of the above probability factors for the variables
obtained, along with whatever probabilities are necessary to account for
the systematic errors involved in real measurements of jet 4-momenta.

  What real $p \bar{p}$ collider events would be candidates for this channel?
Ideally, for a perfectly measured event containing a high transverse
energy lepton, considerable missing transverse energy (for a
neutrino) and at least four jets with clearly identified flavor,
the task of checking whether or not the top production hypothesis
is likely to be correct is straightforward. The likelihood for that event,
with a specific identification of jet flavors, would be
achieved by simply evaluating the {\it a priori\/} probability for the
given
kinematic configuration and applying Bayesian statistics (this was
feasible with the di-lepton event, where there were few
combinations of jets possible[3,6]).
     For real data, however, there are two complications that could
have considerable impact: the sizeable uncertainties in jet energy
measurements and the lack of flavor identification for the jets.
The latter shortcoming leads to the necessity for considering all
possible jet combinations in a given event. Will this necessity
wash out any ability to discriminate top events from a ubiquitous
background? It is our expectation that the kinematic restrictions
favoring small net transverse momentum for decay products and requiring
one jet pair to have invariant mass near the W mass, will
enhance the probability for correct jet combinations.

     Each event in any sample of real Tevatron data, or simulated data as
we will consider here, has a set of "measured" lepton and
jet 4-momenta.  The first step in putting the event through the
fitting procedure is to choose a particular combination of lepton
and jets, with a set of energy and rapidity cuts that assure the
clean determination of the topology of the events. A combination of
four jets is tentatively assigned: one to the semi-leptonic
decaying top and three to the hadronically decaying partner. The
three jets, with 4-momenta p$_1$, p$_2$, p$_3$, form the tentative top or
anti-top momentum $\vec t$=$\vec p_1$+$\vec p_2$+$\vec p_3$, with top
"mass" m$_{t}$*$^{2}$=$t^{2}$.

     The lepton, say electron, and the tentative b-jet 4-momenta,
p$_e$ and p$_b$, define a continuum of ellipses for different top masses.
The minimum allowed value of the square of the mass is
\smallskip
\+&&$(M_W^2+2p_e \cdot p_b)(m_b^2 + 2 p_e \cdot p_b)/2 p_e \cdot p_b$\cr
\smallskip
\noindent
so m$_{t}$*$^{2}$ must be greater than this for a match to be possible. If
m$_{t}$*
can define one of the ellipses, the next step is to determine the
total transverse momentum of the t, $\vec {t}_T$, with a point chosen on the
transverse projection of the ellipse. To choose such a point the
ellipse is parameterized by an uniform angular variable, the angles
are discretized in 5$^o$ units, to be specific, and each corresponding
point on the ellipse will be in consideration.
     Choose one point on the ellipse, as illustrated in Fig.1.
Graphically the vector -$\vec {t}_T$ should lie within 0.1m$_{t}$* of the
given point on the transverse ellipse. If that condition is satisfied the
kinematic configuration is a match. Next move to another angular
point on the ellipse. For every point satisfying this transverse
momentum condition, a match is defined. Usually then, there will be
an arc of matches on the ellipse for a given m$_{t}$*. Each match is
assigned an {\it a priori\/} probability according to its overall kinematics.
The probabilities for all these matches are summed to make one
entry in the probability distribution for this particular t vector,
mass and ellipse.

Two of the jets
must be from W decay for real top decay, so the invariant mass of
at least one pair should lie in the region of the W mass. However, the
spread in pair mass values estimated from Monte Carlo simulations of W decays
in CDF gives a broad distribution, from 40 to 150 GeV, because of
uncertainties in determining jet energies. If the measured pair mass can
be brought to the W mass while staying within the expected measurement
uncertainties, the pair will be a candidate for a W decay. Fixing the pair
mass to $M_W$ provides a strong constraint, as we will see.

     How are the uncertainties in real measurements to be accounted
for in this fitting scheme? The measured value of a jet energy
represents one of a continuum of possible "true" values, with
frequency of occurence for any particular energy given by an empirical
probability distribution. Jet measurement distributions have
been studied carefully by the CDF group with a resulting algorithm
for evaluating the width, $\sigma$, of a roughly gaussian distribution of
energies[10]. (The uncertainty in the lepton measurement is quite
small compared to the jets and will be ignored here.) So to account
for the distribution in a single jet 4-momentum, the $\sigma$ for that
energy is determined. A fixed number, say N, of discrete
weighted values are chosen for the energy varying over
E(measured)$\pm$ a constant number of $\sigma$'s. Since we are selecting
relatively high
energy jets, and directional measurements are better than energy
measurements at CDF, we assume the variation in 3-momentum will be
in magnitude rather than direction and will follow the energy
variation (jet mass being ignored).

     There are four jets in the particular combination we are
considering at this point. In the parameterization just described,
each jet's 4-momentum (and gaussian measurement weight) would vary over
N values as their four energies vary independently over a
hypercubic (N)$^4$ lattice. Thus there would be (N)$^4$ kinematic
configurations for this combination of jets. Because of the constraint
that one pair must have an invariant mass equal to $M_W$ or 80.6 GeV,
the variation of
the pair's energies can not be independent; an independent variation
of each jet's energy in the pair could force the invariant mass farther from
the W mass,
and make the correct kinematic reconstruction more problematic.
The independent variation (over several standard deviations)
of the two energies, say $E_1$ and $E_2$, would fill a rectangular area in
the plane
spanned by the two variable energy values. But each pair of energies in
that plane corresponds to a different invariant mass for the pair.
Forcing the two energies to give the W mass ($E_1 \cdot E_2$ is
proportional to the pair mass) thereby
constrains the energies to lie on a section of an hyperbola. The
uniform division of that hyperbola into N equal "angles" (in the
parameterization through hyperbolic functions) provides the proper
analog to the single jet variation. Associated with each pair of
"corrected" energies on the hyperbola is the product of the gaussian
probability
values for the displacement to the corrected value of each energy from its
measured value. That product of gaussian probabilities will be biggest for
a pair that has its measured invariant mass nearest to the W mass.

     With this W mass constraint, and its parameterization, there
are (N)$^3$ values over which the particular combination of jets'
kinematic configuration can vary.
     Each kinematic configuration is submitted to the same fitting
procedure, matches are obtained and the {\it a priori\/} probability
for that mass m$_{t}$*
obtained. When all the (N)$^3$ configurations are processed a
probability distribution versus top mass for that jet combination is
obtained. It is a measure of the probability that that combination fits the
top hypothesis for each mass.

  For a given event, which
one of the combinations is the "correct" one? In the simulated events,
to be discussed, that is known, of course, but for real data either a
criterion must
be set for selecting a subset or all combinations should be treated
equally. For different combinations of jets,
including the "real" decay fragments, the correctly assigned jet
combination would be expected to have the highest integrated probability.
That suggests
a procedure in which all jet combinations that pass the constrained
fitting criteria be accumulated in the probability for that event.
While wrong choices will contribute to this accumulated distribution, the
correct choice usually should dominate. Furthermore, the relative
probability for each combination is a meaningful quantity, since each
kinematic configuration is treated uniformly with the same
parameterization.

So the distribution with mass of the sum of the
probabilities for the different
combinations in an event is a distribution of probabilistically weighted
"events". How are other events to be treated?
Since there is no way to determine which combination is correct, there is
no distinction between a combination formed from one event or any other.
This leads to adopting a procedure in which all combinations of jets from
all events are
treated uniformly. Any kinematic configuration with a non-vanishing
probabilistic weight at a given mass should be counted in the
distribution at that mass. Hence all the individual weighted combinations
should be added together to form the "accumulated" probability
distribution as a function of the hypothetical top quark mass. This
{\it accumulated probability distribution\/} is just the sum of the probability
distributions from each event. This distribution will provide the means to
separate "top" events from non-top events.

To test the fitting procedure advocated here, we have used several sets
of simulated data, belonging to two categories. Simulations of top pair
production events, with
different levels of complications due to real detectors, were generated and
will be referred to as TTbar sets. The simplest of these, a "toy model", is
a parton level
scheme in which the top pair is produced back-to-back in the center
-of-mass of the incoming quark-antiquark or gluon pair. The energies and
angles of the $t \bar{t}$ are chosen at random as are the directions of
their decay products (through the physical W) in their respective rest
frames, all subject to
correct kinematic constraints. A more complex set will also be used for
TTbar, below.

A second category
consists of Standard
Model Monte Carlo simulations that produce jets and a W that decays
leptonically. These are
the W+3jet and W+4jet sample. Both categories are generated by
starting with the relevent tree level Feynman diagrams. For $t \bar{t}$
production the $q \bar{q}$ annihilation or gluon fusion amplitudes are
relevent. For competing non-top mechanisms, processes in which
quarks or gluons produce a W boson and four hard gluons, or two
gluons and a light quark pair, are initiators[7]. Then the outgoing
partons must fragment into hadron jets through some standard
iteration scheme[9]. Finally the measurement of the kinematics of
those jets and leptons depends on the detection system and the various jet
finding procedures and fragmentation codes. We have
used the CDF detector simulation code QFL, which incorporates the
efficiencies and peculiarities of that particular system as
understood in the 1988-1989 run.

To begin with we show several probability distributions in Fig.2 for
different jet combinations in a single $t \bar{t}$ simulated toy event for
$m_t$=140 GeV.
The probabilities are added together to obtain a probability distribution
for that event, as we proposed above. Note that the peak near 140 GeV is
due to the "correct" combination and dominates over the contributions from
"wrong" combinations, as anticipated.

 When 100 tightly constrained toy events are randomly
generated and put through the fitting procedure the resulting
accumulated probability distribution, Fig.3a, is sharply peaked, with smaller,
insignificant structures arising from wrong jet combinations, as
expected. The broadening of the peak is from the gaussian
probabilities for the CDF indeterminancy of jet energies.

     To simulate the mismeasurements arising from the detection
efficiencies as well as smearing through fragmentation and soft
gluon bremsstrahlung, we randomly alter the values of the jets'
4-momenta using the same CDF determination of the standard
deviation in energy determination and the corresponding inverse
gaussian or error function to reproduce the gaussian distribution
of energy values. The likelihood distribution for these "smeared"
events, Fig.3b, remains peaked at the fixed mass of 140 GeV,
although there is some broadening and the wrong combinations are
relatively more significant. Nevertheless, it remains quite clear
that the identification of the top peak, as well as the mass
determination, are quite striking in the accumulated distribution function.
To be more realistic, cuts were applied: i) three jets
originating from the "hadronic" top decay
were required each to have high transverse momentum; ii) all
four jets were required to have their pseudo-rapidities limited;
iii) the missing transverse energy in the
entire event
was required to be large. These cuts are motivated by
Monte Carlo studies and the general characteristics of existing $p \bar{p}$
detectors, with CDF as an example.

To include the detector effects a full simulation of the CDF detector has
to be performed in the environment of jets of hadrons. A distribution for a
sample of ISAJET+QFL events[9], with the cuts indicated above, was
generated with m$_t$=130 GeV, and
is shown in Figure 4.
Again, the distribution
exhibits a sharp, somewhat displaced peak.

     What of the background for the lepton and jets ? Simulation of
the expected Standard Model contribution to the relevant final
states has been constructed through the mechanism of partons
producing a W boson along with hard gluons and/or quark-anti-quark
pairs[7]. Applying the CDF detector constraints to pseudo-rapidity
and transverse energy of the parton level processes resulted in a
prediction that the background would be no more than roughly 50\% of
the signal[7]. But will that be sufficient to muddy the clean
determination of the toy model by enhancing particular regions of
the three jet mass ? To test that possibility we took a large sample
of the W + 3 jets and W + 4 jets generated[11] with VECBOS Monte
Carlo program
which implements the QCD calculation of Berends, et.al.[7].
A sample of W+4jet events corresponds to an integrated
luminosity of 112 pb$^{-1}$ and that of W+3jet events to 128 pb$^{-1}$.
The events were subjected to the same analysis as $t \bar{t}$ events.
The resulting accumulated probability distributions as a function of $m_t$ and
normalized to the luminosity of 4 pb$^{-1}$ (integrated luminosity of
1988-1989 CDF run)
are shown in Figure 5. The likelihood distribution for $t \bar{t}$ ISAJET+QFL
Monte Carlo events,
normalized to the same luminosity, has
been superimposed on the W+jets' graphs. There is very little background left
above m$_{t}$=120
GeV. The completely different forms of the probability distributions for
TTbar simulations versus W+jets demonstrates that the accumulated
probability distribution function
for {\it real data \/}will effectively distinguish $t \bar{t}$ production from
Standard Model background.

To summarize our results, first note that the
geometric method offers many advantages over a multi-
dimensional, multiply constrained fit of kinematic variables. The
parameterization for a single ellipse is one-dimensional, uniform
and continuous. So the corresponding linear phase space density,
which is also fairly uniform, is well represented by a uniform grid
(set by dividing the ellipse into equal angular steps). The uniform
parameterization is the same for all ellipses, all top masses and
all events. Hence the probability assignment for a given mass in a
given event is commensurate with the probability assignment for any
other mass in any other event. The events can be combined to form
an accumulated distribution with the relative probability of
each event being a meaningful quantity.
     The continuity of the ellipses provides for the unambiguous
determination of kinematical configurations, without the two-fold
ambiguity in solving algebraic constraint equations that can lead to
spurious solutions in handling large samples. The single
parameter ellipse is the geometric realization of the continuum of
solutions to constraint equations.

     From a practical point of view, the ellipse provides an
efficient, systematic and unambiguous realization of the kinematic
fitting procedure. Because there is a single parameter for each
mass, the probability distribution will be smoothly varying with systematic
variation of the measured jet energies. There is no possibility of
falling into local extrema in multi-dimensional phase space, as
there will be in the traditional constrained fit approach.

  Finally we note that the accumulated probability distribution as a
function of mass is completely consistent with the requirements for an
efficient discriminator. The peaks for top events are where they
should be - masses are well determined. The Standard Model background
peaks at low mass, giving virtually no substantial
contribution in the region above 120 GeV. If real events contained a top
contribution in the mass range above 120 GeV (as suggested in the dilepton
event), this procedure would provide a dramatic demonstration of the
existence of such a top quark.
\bigskip
\center { ACKNOWLEDGEMENTS}

Two of us (G.R.G. and K.S.) acknowledge the U.S. Department of Energy for
partial support during the course of this research.
\np
\bigskip
\center {  REFERENCES}

1. For a recent review see G.L.Kane, "Top Quark Physics", University of
Michigan preprint, UM-TH-91-32 (1991).

2. K. Sliwa (CDF Collaboration), in Z$^{o}$ Physics, Proc. 25th Rencontre de
Moriond, Les Arcs, France, 1990, edited by J.Tran Thanh Van (Editions
Frontieres, Gif-s,r-Yvette, 1990), p.459; F. Abe, et.al., Phys.Rev. Letters
68, 447 (1992); P.Langacker
and M.Luo, Phys.Rev. D44,817 (1991).

3. R.H.Dalitz and Gary R. Goldstein, Phys.Rev. D45, 1531 (1992).

4. I.Bigi, Phys.Lett. B175, 233 (1986); I.Bigi, et.al., Phys. Lett. B181,
 157 (1986).

5. F. Abe et.al. (CDF Collaboration) Phys.Rev.Letters 64 (1990) 147;
"Results from Hadron Colliders", Lee Pondrom, in Proceedings of XXV-th
International Conference on High Energy Physics, Singapore 1990; "Search
for top quark at Fermilab Collider", K. Sliwa, in Proceedings of 4-th
Heavy Flavour Conference, Orsay 1991.

6. R.H.Dalitz and Gary R. Goldstein, "The analysis of top-antitop
production and dilepton decay events and the top quark mass",
University of Oxford, Department of Physics - Theoretical
Physics, preprint (1992).

7. F.A.Berends, H.Kuijf, B.Tausk, W.T.Giele, Nucl.Phys.B357, 32
   (1991).

8. F. Abe et.al. (CDF Collaboration), Phys.Rev.Lett. 64, 157 (1990); P.
Bagnaia et.al. (UA2 Collaboration), Phys.Lett. 144B, 283 (1984).

9. ISAJET is described in F.Paige and S.Protopopescu, in "Proceedings of
the Summer Study on the Physics of the SSC, Snowmass, Colorado, 1986,
edited by R.Donaldson and J.Marx (DPF,American Physical Society, New York,
1986), p.320. QFL is described in C.Newman-Holmes and J.Freeman, in Proc.
of the Workshop on Detector Simulation for the SSC, Argonne, 1987, ed. L.
Price (A.N.L. Report no. ANL-HEP-CP-80-51),pp.190,285.

10. CDF Collaboration, F. Abe et al, Phys. Rev. Lett. 68 (1992) 1104; we
appreciate the help of Naor Wainer for providing us with the subroutine which
incorporates all the information on jet errors.

11. We appreciate the work of Jose Benlloch in generating the VECBOS samples.

\vskip 2.0 truecm
                         FIGURE CAPTIONS

1. The vector and ellipse construction for a possible match of
transverse momentum for a simulated top pair production event.

2. Probability distributions for three individual events, generated with
M$_t$=140 GeV. Probabilities for ALL combinations for each single event are
added together to form a single distribution for a single event.

3. Accumulated probability distributions for a sample of 100 events generated
with
M$_t$=140 GeV without (a) and with (b) smearing of the jets according to
standard CDF parametrization of the error on the jet energy measurement.

4. Accumulated probability distributon for a sample of 100 ISAJET+QFL events
generated
with M$_t$=130 GeV. The mass shift (from 130 GeV) is understood as an artifact
of some incompatibilities in simulation.

5. Accumulated probability distributions for VECBOS W+3jets and W+4jets
samples.

\eject
\end